\documentstyle[12pt,aasms4,flushrt]{article} 
\def\lsim{\lower0.6ex\vbox{\hbox{$ \buildrel{\textstyle 
<}\over{\sim}\ $}}} 
\def\gsim{\lower0.6ex\vbox{\hbox{$ \buildrel{\textstyle 
>}\over{\sim}\ $}}} 
\begin{document} 
 
%\twocolumn 
 
\title{Temperature of the Central Stars of Planetary Nebulae and the Effect 
of the Nebular Optical Depth} 
 
\author{R. Gruenwald and S. M. Viegas} 
\authoraddr{Instituto Astron\^omico e Geofisico, USP, Brazil\\} 
 
\vspace{.2in} 
 
\begin{abstract} 
The effect of the nebula optical depth on the determination of the  
temperature (T$_*$) of the central stars in planetary nebulae is discussed. 
Based on photoionization models for planetary nebulae with different optical 
depths, we show, quantitatively, that the details of the distribution of 
the H and He II Zanstra temperatures are mainly explained by an optical 
depth effect; in particular, that the discrepancy is larger for 
low stellar temperatures. The results also show that for high stellar 
temperatures the He II Zanstra temperature underestimates the stellar 
temperature, even for high optical depths.
The stellar temperature, as well as the optical depth, can be obtained from 
a Zanstra temperature ratio (ZR) plot ZR = T$_Z$(He II)/T$_Z$(H) versus 
T$_Z$(He II).
The effects of departures from a blackbody spectrum, as
well as of the He abundance in the nebulae, are also discussed. 
For nebulae 
of very low optical depth and/or high stellar temperature the distribution 
ZR versus T$_Z$(He II) only provides lower limits for T$_*$.
In order to obtain better values for the optical depth and T$_*$,
we propose the use of the line intensity ratio He II/He I versus
T$_Z$(He II) diagram.
\end{abstract} 

 
\section{Introduction} 
 
The temperature of the central stars of planetary nebulae (PNs) 
is an essential parameter for evolutionary  
studies as well as for an analysis of the nebulae themselves.  
The different methods for the determination of the stellar temperature 
and the   
problems involved were extensively discussed by Pottasch (1984)  
and Kaler (1985a, 1989). 
The most common method was
suggested by Zanstra (1931) and further developed by  
Harman \& Seaton (1966). The temperature of the  
ionizing star of a planetary nebula is calculated from the ratio  
between the flux of a recombination line 
and the stellar continuum flux at a given frequency.   
The Zanstra method assumes that all photons above the H (or the He$^+$)  
Lyman limit are absorbed within the nebula and that each recombination  
eventually results in a Balmer photon. The total ionizing flux can then be  
related to the total flux of a recombination line.  
The Zanstra method yields then two values for the stellar  
temperature: the Zanstra temperature obtained from the intensity of a  
hydrogen recombination line, T$_Z$(H),  
and the Zanstra temperature obtained from a He II recombination line,  
T$_Z$(He II). 
When applied to observed nebulae,  T$_Z$(H) is generally lower than  
T$_Z$(He II); the difference can reach values of the order of 60,000 K 
(Kaler 1983b), and the temperature ratio can reach a factor 
higher than 3 (Kaler 1983b, 1985a). This is the well known Zanstra 
discrepancy, which is  
 stronger for PNs with low stellar  
temperatures. In fact, many PNs with  T$_Z$(H) $<$ 100,000 K have  
T$_Z$(He II) $>$ T$_Z$(H)  
while, for higher stellar temperatures, both Zanstra temperatures are similar  
(Pottasch, 1984; Gathier \& Pottasch 1988, 1989).  
Another important point is that the calculated Zanstra temperatures  
do not reproduce the high temperatures predicted by evolutionary models  
(Kaler 1985a, 1989; Stasi\'nska \& Tylenda 1986). 
 
Another interesting feature that could be related to the 
Zanstra temperature issues is the distribution of objects  
in a log L - log T plane. As noted by Shaw \& Kaler (1989), when this  
distribution is based on Zanstra temperatures there is a dense crowd of  
planetary nebula nuclei with temperatures of $\sim$ 100,000 K and a 
strong decrease towards higher temperatures, what these authors call the  
``Zanstra wall''. 
 
The causes of the Zanstra  discrepancy have been  
discussed by several authors (Pottasch 1984; Kaler 1985a, 1989; 
Henry \& Shipman 1986;Stasi\'nska \& Tylenda 1986; 
Kudritzki \& M\'endez 1989; Gabler, 
Kudritzki \& M\'endez 1991; M\'endez, Kudritzki \& Herrero 1992) and can be 
related to  the following effects:  
(1) optical effects, i.e., nebulae exhibiting a Zanstra discrepancy would  
be optically thin to photons ionizing H yet optically thick to those  
ionizing He$^+$; 
(2) differential dust absorption in the nebula; 
(3) the stellar continuum differing from the usually  assumed blackbody  
spectrum; for example,  
an excess of photons with energies beyond the He$^+$ ionization potential  
would result in a high He II Zanstra temperature. 
 
Objects with a large discrepancy show fainter low-ionization lines,  
suggesting that the effect 1 is the correct interpretation (Kaler 1983b).  
Simple evolutionary models of planetary nebulae predict  
T$_Z$(H) $\neq$  T$_Z$(He II) during the nebula lifetime because of the  
variation of their optical depth (Tylenda et al. 1994; see also 
Sch\" onberner  
\& Tylenda 1990). 
The effect of dust on the calculated Zanstra temperature is discussed by  
Helfer et al. (1981). For an assumed absorption law  
(with a high opacity around 50 eV), Zanstra temperatures underestimate  
the actual stellar temperature. The effect is more intense for T$_Z$(H) 
and leads to the Zanstra discrepancy. However,  
dust appears to be only important in some specific nebulae (Kaler 1985a).  
Departures from the blackbody, in particular an excess of photons  
beyond the He$^+$ ionization potential, are suggested  
by observational studies of  many central stars of planetary  
nebulae (Kaler 1985a;  
see also references in Henry \& Shipman 1986). An ionizing spectrum  
with an excess   
of high energy photons (relative to a blackbody),  
produced by a star with a less than solar atmospheric  
He abundance, leads to a He II  
Zanstra temperature higher than T$_Z$(H) (Henry \& Shipman 1986). 
 
Detailed theoretical analyses, using photoionization models and applied  
to optically {\it thick} nebulae, are presented by Stasi\'nska \& 
Tylenda (1986) and Henry \& Shipman (1986). 
Stasi\'nska \& Tylenda (1986) show that for low stellar temperatures  
(T$_*$ $\leq$ 100,000 K), both Zanstra temperatures are similar.  
For higher stellar temperatures, 
T$_Z$(HI) is larger and T$_Z$(He II) is lower than T$_*$. However,  
this result is  opposite to what is obtained from observations. 
Analyzing only models with T$_*$ $\leq$ 150,000 K, Henry and Shipman  
(1986) conclude that T$_Z$(H) is a good measure of the stellar  
temperature. In fact, for these values of T$_*$, 
T$_Z$(H) and T$_Z$(He II) are similar (Stasi\'nska \& Tylenda 1986) 
and provide a good stellar temperature estimation {\it if the nebula 
is completely optically thick}.  
 
Other methods for determining the stellar  
temperature include modelling of stellar absorption line profiles,  
ionic ratios, fitting of model atmospheres, the energy balance or Stoy 
method, stellar UV energy distribution, etc. The results from these 
different methods were compared  
with the values given by the Zanstra  
method in order to explain the Zanstra discrepancy; however,  
these methods give discordant results and have many uncertainties (see, for 
instance, Kaler 1985a, Stasi\'nska \& Tylenda 1986, and Kaler 1989). 
 
Many authors have adopted T$_Z$(He II) as representative of the stellar  
temperature assuming that the Zanstra discrepancy is due to  
an optical depth effect 
(Kaler 1983b; Gleizes, Acker \& Stenholm 1989; Kaler, Shaw \& 
Kwitter 1990; Kaler \& Jacoby 1991; Stanghellini, Corradi \& Schwarz 1993). 
However, T$_Z$(He II) may not be a good indicator of the stellar temperature  
since it never reaches values as high as predicted by theoretical stellar 
evolutionary studies. 
 
Usually the discussion of the Zanstra discrepancy  
in terms of an optical depth effect is based on  
 optically thin {\it or} thick  
nebulae at 13.6 eV or 54.4 eV. It is also implicitly assumed 
that the nebula is optically thick at 54.4 eV at a distance to the 
central star smaller than that corresponding to 13.6 eV. 
Harman \& Seaton (1966) suggest the following criteria for the complete 
absorption of H$^0$, He$^0$, and He$^+$ ionizing-photons: presence of [OI] 
lines, He I images smaller than H I images, and He$^{++}$ fractional 
abundance $\leq$ 0.75, respectively. 
For Pottasch (1984) the Zanstra method must work for  
$\tau_{13.6}$ $>$ 1.  
Some authors define a criterion to distinguish between optically thin and  
thick objects using the ratio of the Zanstra temperatures, 
ZR = T$_Z$(He II)/T$_Z$(H). 
For example, for Shaw \& Kaler (1985)  
the nebula is optically thick to the H Lyman continuum   
when ZR $\lsim$ 1.2,  
while for ZR $\gsim$ 2.5 and  
He II $\lambda$4686/H$\beta$ $\gsim$ 0.9 it is thin  
for He$^+$ Lyman continuum photons, T$_Z$(He II) being a lower limit for  
the stellar temperature. The criterion used by Kaler \& Jacoby (1989), 
based on line intensities of low-ionization lines, states that a 
nebula is thick when  
[O II] $\lambda$3727/H$\beta$ $\geq$ 1 and 
[N II] $\lambda$6584/H$\alpha$ $\geq$ 1. 
However, planetary nebulae can present a large range of  
optical depths, depending on the quantity of matter. Furthermore, 
even considering the central stellar radiation as a  
blackbody, different stellar temperatures correspond to different 
ratios between the number of ionizing photons with energy higher 
than 54.4 eV and those higher than 13.6 eV. Thus, the radial ionic 
distribution for 
H and He varies with T$_*$, changing the relative sizes of the H$^+$ and
He$^{++}$ zones with the nebula optical depth at 13.6 eV and 54.4 eV. 
As remarked by Stasi\'nska \&  
Tylenda (1986), the radiation transfer is much more complicated  
than assumed by the Zanstra method. 
 
In brief, the Zanstra temperature is commonly used in  
the literature for planetary nebula modelizations as well as evolutionary  
analysis, and low optical depth must be at least a partial  
explanation for the Zanstra discrepancy. However, a more 
detailed analysis is required in order to explain the issues 
listed above. In this paper, 
a careful analysis of the effect of the nebula optical depth  
on the determination of the Zanstra temperatures is intended.  
The effects on the Zanstra temperatures due to 
deviations of a blackbody spectrum and due  
to an overabundance of He in the nebula are also discussed. The  
theoretical models used in our analysis are  
described in \S 2. The results for the Zanstra temperature ratio (ZR) 
and its behavior  
with the stellar temperature and with the nebula optical depth  
appear in \S 3, which also includes a comparison with values  
derived from PNs observations. An alternative method  
to estimate the  
temperature of central stars of planetary nebulae is suggested and  
discussed in \S 4. The conclusions are outlined in \S 5. 
 
\section{Planetary nebulae models and theoretical values for the Zanstra 
temperatures} 
 
Models for typical planetary nebulae are generated with the photoionization 
code AANGABA (Gruenwald \& Viegas 1992). 
The physical conditions of the gas are determined by solving the coupled  
equations of ionization and thermal balance for a spherical symmetric cloud. 
Several processes of ionization and recombination, as well as of gas heating  
and cooling, are taken into account. The transfer of the primary and diffuse  
radiation fields is treated in the ``outward-only'' approximation.  
For the radiation-bounded models (equivalent to completely optically thick 
nebulae), the calculations stop when the fractional abundance H$^+$/H 
reaches 10$^{-4}$, defining the maximum radius for the ionized nebula, 
R$_{max}$. Matter-bounded 
models (with a nebula radius less than R$_{max}$) will also be discussed.
The input parameters are the ionizing radiation spectrum, the gas density, and 
the chemical abundance for the elements included in the calculations  
(H, He, C, N, O, Ne, Mg, Si, S, Cl, Ar, and Fe). 
A range of input parameters, typical of planetary nebulae (Pottasch 1984), 
is assumed: T $\geq$ 50,000K, L$_*$ = 30 - 20,000 L$_{\odot}$ and  
n$_H$ = 10$^2$ - 10$^6$ cm$^{-3}$.  
In order to discuss  
the Zanstra temperature for very hot stars, which are predicted by  
evolutionary models, a maximum stellar temperature of 500,000 K is adopted. 
A blackbody spectrum is assumed for the ionizing radiation, but
the effects due to departures from this kind of spectrum will  
also be discussed. Concerning the chemical abundances, average  
values for planetary nebulae, as given by Kingsburgh \& Barlow (1994), are 
assumed. 
For elements not given by these authors, the solar value is 
adopted (Grevesse \& Anders 1989).  
 
The He II $\lambda$4686  
and H$\beta$ fluxes obtained for the theoretical nebulae are used to 
derive  T$_Z$(H) and T$_Z$(He II) by the standard Zanstra method. 
For each set of input parameters,  T$_Z$(H) and T$_Z$(He II) are 
calculated for different values of the nebula optical depth at the 
H Lyman limit. 
 
\section{Theoretical versus ``observed'' Zanstra temperatures} 
 
In the following section, the assumed energy distribution of the  
ionizing radiation is fixed  
(blackbody). Thus, any discrepancy between the   
temperature adopted for the central  
star, T$_*$, and the derived Zanstra temperatures is not due to  
the assumed spectrum but inherent to the method.  
 
Our results show that T$_Z$(H) reproduces fairly well the stellar  
temperature for {\it optically thick nebulae} ionized by a star  
with T$_*$ $<$ 150,000 K, 
in agreement with  Henry \& Shipman (1986) and Stasi\'nska \& Tylenda (1986).  
For higher stellar temperatures, T$_Z$(H) is greater than T$_*$, and
the difference between these two values increases with T$_*$. 
These results agree with those of  
Stasi\'nska \& Tylenda (1986). We find, however, that  
the deviation of T$_Z$(H) relative to T$_*$ is slightly smaller. 
As already discussed by Stasi\'nska \& Tylenda (1986) 
the deviation of Zanstra temperatures from the stellar temperature is due   
to the fact that each He$^{++}$ recombination gives more than one 
photon that ionizes H, and the proportion of photons 
ionizing  He$^+$ increases with the stellar temperature. Furthermore,  
a fraction of the high energy photons are in fact absorbed  
by H and not by He$^+$. 
A detailed discussion on the generation of H ionizing photons following  
the He$^+$ and He$^{++}$ recombination can be seen in Osterbrock (1989). 
 
\subsection{Emitting volumes of H$^+$, He$^+$, and He$^{++}$}
 
Before discussing the influence of the nebula optical depth on the derived 
Zanstra temperatures, it is useful to illustrate how the relative sizes of 
the H$^+$, He$^+$, He$^{++}$ Str\" omgren spheres change as a function of 
the stellar temperature.
The variation of the fractional abundances of H and He ions 
with the position in the nebula is shown 
in terms of r/R$_{max}$ in Figure 1a (left panels)
where r is the distance from the center of 
the nebula and R$_{max}$ is the maximum dimension of the ionized region 
(see \S2).
The results given in Figure 1 correspond to models with L$_*$ = 3000 
$L_\odot$ and n$_H$ = 10$^4$ cm$^{-3}$. For low stellar temperatures the  
He$^{++}$ Str\" omgren radius, R$_{He^{++}}$, is much smaller  
than R$_{He^{+}}$ or R$_{H^{+}}$, as expected. However, as T$_*$ increases, 
R$_{He^{++}}$ approaches R$_{H^{+}}$. 

For a matter-bounded nebula (with a total extent less than R$_{max}$) the 
emitting zones of H$^+$, He$^+$, and He$^{++}$
can be smaller than their corresponding Str\" omgren spheres. In this case, 
the emitted line intensities will be lower than those emitted by a 
radiation-bounded nebula. For a given reduction of the nebula extent, the 
size of the emitting zones of each of these ions will be differently 
affected, depending on the temperature of the central star. 
For nebulae with low 
stellar temperatures, a reduction of the nebula size affects mainly the H$^0$, 
H$^+$, and He$^0$ zones. Thus, the smaller the nebula radius, the lower 
T$_Z$(H), 
while T$_Z$(He II) may still be a good indicator of the stellar temperature.
For increasing stellar temperatures, 
the volumes of the H$^+$ and He$^{++}$ zones tend to be equal (Fig. 1a). In 
this 
case a reduction of the nebula size can result in a matter-bounded nebula 
where the  H$^+$ and He$^{++}$ zones are almost equally  
affected. In this case, both T$_Z$(H) and  T$_Z$(He II) underestimate 
the stellar temperature. 

\subsection {The effect of the nebula optical depth}

The possible underestimate of the stellar temperature, due to the fact that 
a nebula may not be radiation-bounded, can be discussed as an optical 
depth effect.  
If the nebula has not enough material to be radiation-bounded, its radius is 
smaller than R$_{max}$, and the nebula 
optical depth at a given frequency will be lower than the optical depth  
of a radiation-bounded nebula. 
The reduction of the nebula radius (creating a matter-bounded nebula)
will differently affect the nebula optical depth of the 
H$^0$, He$^0$, and He$^+$ continua.
The fractional ionic distribution of H and He ions with the optical depth at 
the H Lyman limit ($\tau_{13.6}$) is shown in Figure 1b. 
As seen in \S 3.1, for a radiation-bounded nebula with low
T$_*$ the He$^{++}$ Str\" omgren radius is much smaller than the 
H$^+$ Str\" omgren radius.
Thus matter-bounded nebulae can have an optical 
depth at the H Lyman continuum, $\tau_{13.6}$, close 
to unity, while the optical depth at 
the He$^{++}$ Lyman limit, $\tau_{54.4}$, is much higher.
The object is then optically thin to the H-ionizing photons
and optically thick to the He$^+$-ionizing photons, leading to ZR higher 
than 1. In this case, T$_Z$(He II) provides a better estimate of the 
stellar temperature. 
As the stellar temperature increases, the 
optical depth at the H and He$^+$ Lyman limits tend to have similar values; 
both will be reduced if the radius of the nebula is smaller than 
that of a radiation-bounded nebula. 
In this case, neither T$_Z$(H) nor T$_Z$ is a 
good indicator of the stellar temperature. 
 
The effect of the nebula optical depth on the derived Zanstra temperatures 
is shown in a ZR versus T$_Z$(He II) plot (Figs. 2a and 2b) 
for the same models as in Figure 1. 
Each solid line corresponds to models with a given stellar temperature; 
the nebula optical depth decreases with increasing ZR. The curves are 
labeled by the stellar temperature in units of 1000 K. 
The dashed curves connect the results of completely 
optically thick models (radiation-bounded nebulae) with different 
stellar temperatures; these results correspond to the minimum ZR 
value for a given stellar temperature. 

Recalling the ionic distribution 
shown in Figures 1a and 1b, the behavior of ZR shown by the curves in 
Figure 2 can be easily understood: 
(1) For T$_*$ $<$ 150,000 K, the He$^{++}$ zone is inside the H$^+$ 
zone and much smaller. Matter-bounded models with decreasing 
$\tau_{13.6}$ would result in weaker H$\beta$ emission line, 
while the He II $\lambda$4686 line is unchanged. Thus, starting 
at the minimum value, corresponding to the optically thick 
model, ZR increases while T$_Z$(He II) is practically constant. 
When the optical depth is low enough to affect the He$^{++}$ zone, 
ZR still increases but T$_Z$(He II) decreases 
and the curves turn to the left; (2) For higher stellar temperatures, 
the decrease of T$_Z$(He II) with $\tau_{13.6}$  happens closer 
to the optically thick value (the volumes of the H$^+$ and He$^{++}$ 
zones are similar) and ZR increases slowly. 
In each of the solid lines in Figure 2a, the points corresponding to 
$\tau_{13.6}$ = 1, $\tau_{13.6}$ = 10, and $\tau_{54.4}$ = 1 are
indicated, respectively, by crosses, triangles, and dots. 
Notice that for 
T$_*$ $>$ 200,000 K, a nebula can be optically thin for He$^+$-ionizing 
photons, even for ZR $\sim$ 1, though thick for photons above the H 
Lyman limit. 
 
The theoretical results  also show that the Zanstra method tends 
to underestimate the stellar temperature. The effect 
is larger for higher stellar temperatures, 
even for high optical depths. This may explain the "Zanstra wall" 
in the log L - log T plot, since high-temperature stars, 
predicted by stellar evolutionary models, are penalized by the  
Zanstra method. 
 
\subsection {Confronting the theoretical results with the observations}

Values for ZR and T$_Z$(He II) derived from observations for a large sample 
of PNs are plotted in Figure 2b in order to be compared to the theoretical 
results.
For each object several values of the Zanstra temperatures 
can be found in the literature. The criteria 
used to select the objects and the values of the Zanstra temperatures 
plotted in Figure 2b are 
the following: (1) if the Zanstra temperatures coming from different 
authors are similar (difference less than 20$\%$ from the average value), 
their average value is taken; (2) if the same author presents discordant data 
for the same object, the more recent value is taken; 
(3) if all the data for a given  object are discordant, the  
object is not included. The values for both Zanstra temperatures 
were taken from: 
Martin (1981); Kaler (1983b);  
Pottasch (1984); Reay et al. (1984); Shaw \& Kaler 1985; Viadana \& de Freitas 
Pacheco (1985); de Freitas Pacheco, 
Codina \& Viadana (1986); Gathier \& Pottasch (1988, 1989);  
Gleizes at al. (1989); Jacoby \&  
Kaler (1989); Shaw \& Kaler (1989); Kaler et al. (1990); Kaler \&  
Jacoby (1991); M\'endez et al. (1992). 
Since  we are discussing the standard Zanstra method,  
Zanstra temperatures corrected by the Stasi\'nska-Tylenda  
effect (1986) were not included. 
 
Most of the observational points for ZR and T$_Z$(He II) are inside  
the region defined by the theoretical curves that correspond to  
ionizing stars with a blackbody spectrum. 
Our results naturally explain the trend shown by the 
observational  
values: for lower H Zanstra temperatures ($\leq$  100,000 K)  
many planetaries may have  T$_Z$(He II) $>$ T$_Z$(H),  
i.e., ZR $>$ 1, while for higher  
temperatures the difference between these temperatures is smaller. 
Such a behavior, referred to as ``strange'' by Pottasch (1984),  
induced Gathier \& Pottasch (1988) to discard the optical depth explanation,  
since nebulae with higher stellar temperature should be older and  
optically thinner. 
The distribution of ZR versus  T$_Z$(H)
presents a similar trend. Notice that a decreasing ZR ratio  
with increasing T$_Z$(H) was obtained by Gathier \& Pottasch (1988)  
with a sample including fewer objects. 
 
The variation of the stellar temperatures with the optical depth in Figure 2 
can solve some problems raised in the literature. One such problem is the 
temperature of the ionizing star of NGC 1360. The Zanstra temperatures for 
this object (34,900 K and 79,300 K, from the references given above) are much 
smaller than the temperature obtained from UV measurements (100,000 K; 
Pottasch et al. 1978). In Figure 2b the position of this 
object is in the region where the lines are crowded; the nebula is thus 
optically thin and the star can have a higher temperature than that given
by T$_Z$(He II), as suggested by the UV data. Also, the discussion 
(Kaler \& 
Hartkopf 1981) regarding A43 (a thin and high-excitation nebula with a low 
He II Zanstra temperature star) and A50 (medium excitation, thick and high 
T$_Z$) must be reviewed, since, following our results (Figure 2), 
the central star of A43 
can have a temperature much higher than T$_Z$(He II).

\subsection{Other effects} 
 
In the previous section, using the results from photoionization models 
corresponding to a given value of the stellar luminosity and gas density 
and assuming a blackbody spectrum for the ionizing radiation, it was 
shown that the 
main issues concerning the Zanstra temperatures can be explained 
by an optical depth effect. In the following discussion  
the results for different values for the stellar luminosity and/or  
the gas density, as well as for an ionizing radiation 
spectrum deviating from a blackbody shape, are presented.  
 
First, still adopting a blackbody spectrum, we discuss the results 
corresponding  to the whole range of adopted values for the stellar 
luminosity and gas density. 
We verified that, as long as T$_*$ $<$ 200,000 K, the behavior of  
ZR with T$_Z$(He II) is the same as discussed in \S\S3.1 and 3.2, 
For T$_*$ $\geq$ 200,000 K and $\tau_{13.6}$ $>$ 1, models with low stellar 
luminosities 
($\leq$ 100 L$_{\odot}$) {\it and} low gas density ($<$ 10$^3$ cm$^{-3}$)
can give Zanstra temperatures lower than those obtained with the 
standard models discussed above. The differences between the Zanstra and 
stellar temperatures increase with increasing T$_*$ and decreasing 
values for L$_*$ and n$_H$. For example, for T$_*$ = 300,000 K, a maximum 
difference occurs for $\tau_{13.6}$ $\sim$ 10, L$_*$ = 10 L$_{\odot}$, and  
n$_H$= 100 cm$^{-3}$, when both Zanstra temperatures decrease by 
$\sim$ 20 $\%$, increasing the difference between Zanstra and effective 
stellar temperatures. However, only a few PNs would have such low stellar 
luminosities {\it and} gas densities.
 
A number of authors explain the Zanstra discrepancy by an excess of  
photons with energy above 54.4 eV in the ionizing spectrum. This could 
explain the high values of 
T$_Z$(He II) compared to  T$_Z$(H). As discussed by  
Henry $\&$ Shipman (1986), observations and models imply  
an excess of photons beyond the He$^+$  
threshold in numerous planetary nebula nuclei. Such an excess 
could be produced by a stellar atmosphere with subsolar He abundances 
and would lead to T$_Z$(He II) higher than  T$_Z$(H) when 
compared with models where a  blackbody is assumed.  

To show the effect of a spectrum presenting an excess of high-energy 
photons above 54.4 eV, we discuss the results of photoionization models 
with an ionizing radiation spectrum of a pure H atmosphere 
(Wesemael et al. 1980). For example, 
for a completely optically thick nebula around a 150,000 K star, 
T$_Z$(He II) is 6$\%$ higher and T$_Z$(H) is 13$\%$ lower compared to the 
corresponding blackbody results. For decreasing optical depths, the curves 
tend rapidly to those corresponding to blackbody models. In brief, only
for completely optically thick nebulae an excess of high
energy photons will affect (in a small amount) the calculated Zanstra 
temperatures.
 
\subsection{Outsiders} 
 
Some objects shown in Figure 2b are outside the area covered by 
the models. 
These nebulae can be either above the area limited by the curves 
or below it. Those above the curves limiting the high  
values of ZR could be explained by an error in T$_Z$(He II)  
of the order of 5 - 10 $\%$. 
However, checking more carefully, it can be  
verified that most of these nebulae are Abell nebulae,  
including NGC 246, the prototype of the class (Abell 1966). Observations  
of the central star of some of these nebulae show characteristics of  
high stellar temperatures; furthermore, the nebulae have low surface  
brightness and large angular diameter (Abell 1966).  
These objects are probably in an advanced evolutionary stage. 
Some of them are known to have very  
high nebular He abundance in their inner regions (Jacoby \& Ford 1983;  
Guerrero \& Manchado 1996). 
Calculations for He-rich nebulae show that abundances up to He/H = 0.25 can 
explain the positions of the nebulae lying above the limiting curves 
in Figure 2b. Results for T$_*$ = 150,000 K and He/H = 0.20 are shown in 
Figure 3 by the dot-dashed line.
Note that for high optical depths the curves for the same 
stellar temperature but different He abundance are superposed. For T$_*$ = 
150,000 K, the curves  
separate from each other when $\tau_{13.6}$ $<$  2.5. The reason is 
that increasing the He abundance, the He$^{++}$ zone decreases relative 
to the H$^+$ zone by 
36$\%$ in volume for this value of T$_*$. Thus, a 
decrease of the nebula optical depth will only affect the He$^{++}$ zone 
[and consequently T$_Z$(He II)] when T$_Z$(H) is very low. Thus, the 
area covered by the models stretches toward higher ZR, including the 
high ZR objects. 

Large nebulae have been studied in detail by Kaler and collaborators 
(Kaler 1981, 1983b; Kaler \& Feibelman 1985; Kaler et al. 1990).
For many of these nebulae the color temperature obtained from UV 
observations are well above their Zanstra temperatures. This is consistent 
with our results (Fig. 2b) that show that the stellar temperature can be higher
than T$_Z$(He II).
Because of their large diameters, the density in the Abell 
or other large nebulae may be smaller than the one assumed for the standard 
models. However, as mentioned above, results for the Zanstra temperatures 
with different densities are similar.

The above results do not necessarily mean that all the nebulae lying above 
the limiting curves are He-rich, or, inversely, that all He-rich nebulae 
have positions above the curves plotted in Figure 2. The same questions 
can be asked
regarding the nebulae size. From the 19 nebulae above the curves, 14 have 
calculated or limiting values for the abundance; from these, only one have He 
abundance definitively below the average value for planetary nebulae as given 
by Kingsburgh \& Barlow (1994). However, He-rich nebulae are also found in 
regions of higher optical depth in the diagrams. Regarding the size, all 
nebulae above the curves, except two (Hu 1-2, with 0.018 pc, and Cn 1-2, with 
no value calculated for the radius) have radius larger than 0.15 pc 
(Cahn, Kaler, \& Stanghellini 1992). But large nebulae are spread everywhere 
in the diagrams. 
In brief, large and/or He-rich nebulae can be found in any location on the 
diagram, but most of those above the limiting curves are large and He-rich.  

As discussed below, a He-rich atmosphere may provide the explanation for 
the outsiders with very low ZR.
A high He abundance in the inner parts of a nebula can indicate a  
high He in the upper layers of the star, including their atmospheres.  
The energy spectrum of a He-rich atmosphere will show a deficit  
of high energy (E $>$ 54.4 eV) photons because of the contribution of He  
to the stellar continuum opacity.  
Henry \& Shipman (1986) discarded a high He abundance in the stellar 
atmosphere, since they do not explain the Zanstra discrepancy shown by  
most PNs (ZR $>$ 1), concluding that the atmospheres have subsolar  
He abundance (and an excess of high-energy photons). 
Our results show that the main effect originating the Zanstra discrepancy is   
the optical effect. However, photoionization models assuming an ionizing 
spectrum with a deficit of photons with energy higher than 54.4 eV  
indicate that ZR is less than unity in the optically thick case, lowering the 
limit defined by the models.
Results for models with various stellar temperatures and an ionizing 
spectrum showing a deficit of  
a factor of five in the flux of high-energy photons, relative to a blackbody,
are shown by the dotted lines in Figure 3.
The corresponding blackbody results are shown by the solid lines.
Thus, 
such ionizing spectrum, presenting a deficit of high-energy photons, may 
explain the outsiders with low ZR. 
 
\section {Stellar temperatures and the He II/He I line intensity ratio}
 
The results presented in Figures 2 and 3 can be used to  
obtain the nebula optical depth as well as a value for the stellar 
temperature more accurately than that given by the Zanstra temperatures. 
 However, for optically thin nebulae,  
the curves are crowded up and the temperature is not well defined. 
The same occurs for high stellar temperatures, even at high optical depths. 
Thus, another method for obtaining the stellar temperature is required.
 
Since the optical depth is a major factor of the stellar temperature 
determination,   
line intensity ratios produced by ions in   
different ionization stages can be used to distinguish nebulae with   
different optical depths. When plotted against  
T$_Z$(He II), many of these line ratios also  
show a crowding of the curves corresponding to different models.  
The best ratio discriminating the results for different models is the ratio 
between He II and He I line intensities. The results for 
He II $\lambda$4686/He I $\lambda$5786 and 
He II $\lambda$4686/He I $\lambda$4471 versus T$_Z$(He II) are 
presented in 
Figures 4a and 4b, respectively, for the same models of Figures 1 and 2.
Notice that, for a given range of T$_*$ and $\tau_{13.6}$, 
particularly for low optical depths, the curves in Figure 4 are more widely 
spaced and provide a better determination of these parameters than 
the curves shown in Figure 2. 

Regarding the nebulae in the crowded region of Figure. 2b, for which 
there is an 
uncertainty in the determination of T$_*$, the stellar temperature may be 
obtained from Figure 4. Besides the 19 nebulae above the curves in Figure
2b, 28 
nebulae are in the region where the results for T$_*$ are just lower limits.
For all these nebulae, only 20 have measured intensities for He II and He I 
lines. With 
intensities taken from the literature 
(Torres-Peimbert \& Peimbert 1977; Aller \& Czyzak 1979, 1983; 
Jacoby \& Ford 1983;
Kaler 1983a; Kaler 1985b; Manchado, Mampaso \& Pottasch 
1987; Peimbert \& Torres-Peimbert 1987; Kaler et al. 1990;
Acker et al. 1991; 
de Freitas Pacheco, Maciel \& Costa 1992; Stanghellini, Kaler \& Shaw 1994;  
Kingsburgh \& Barlow 1994), 
the stellar temperatures 
obtained from Figure 4 are higher than T$_Z$(He II) by about 10$\%$ in 
general but can 
reach 32$\%$. 
For the selected nebulae, the maximum observed value for 
He II $\lambda$4686/He I $\lambda$5786
is 1.85dex for NGC 4361 (there is also a measured lower limit for its central 
region of 2.25dex). For He II $\lambda$4686/He I $\lambda$4471 the maximum 
measured ratio is 2.37dex, for 
NGC 2022. Higher ratios, corresponding to  
lower optical depths, are not measured since He I   
is too faint to be detected. Notice that for high T$_*$ ($>$ 150,000 K), 
He I can be very faint even for $\tau_{13.6}$ higher than unity, since, 
as shown 
in Figure 1, the higher T$_*$ the smaller the He$^+$ region. So, for 
nebulae with low ZR and no detected He I line, the stellar temperature 
can be much higher than T$_Z$(He II). 
As long as the He II Zanstra 
temperature is known, this alternative method can be used 
and may be considered as a second-order approximation to the 
stellar temperature, providing values closer to the real  
stellar temperature, even for nebulae showing a high He II/He I line ratio.

\section {Conclusions} 
  
One of the problems concerning the understanding 
of PNs and their evolution is the determination of their stellar temperature.  
The Zanstra method is generally used, although the  
H and He II Zanstra temperatures may be discrepant and may 
 underestimate the stellar temperature. 
Many authors suggested that the dominant mechanism  
explaining the Zanstra discrepancy 
is the optical depth. Because T$_Z$(He II) is less 
affected by the nebula optical depth, these authors 
suggest adopting  T$_Z$(He II) as the true stellar temperature. 
Besides the importance of a good determination of the stellar temperature  
of planetary nebulae for the analysis of the emission-line spectrum and the
chemical abundance determination of PNs, 
let us recall that  
the position of the central star of PN on the H-R diagram is  
a crucial test of evolutionary models from the AGB to the white dwarf  
stages. 
 
Here the nebula optical depth effect is analyzed  in detail using 
photoionization 
models. From the theoretical H$\beta$ and He II $\lambda$4686 line 
intensities, 
the Zanstra temperatures are calculated and compared 
to the adopted stellar temperatures for a variety of models 
with different optical depths.  
Our results show that the nebula optical depth is the main factor explaining  
the behavior of the Zanstra temperatures with the stellar temperature. 
The Zanstra discrepancy mainly occurring for low stellar 
temperatures is clearly explained 
by the changes induced by the optical depth 
in the relative ionic distribution in the nebula (Figs. 1 and 2).
Another consequence is that even the He II Zanstra temperature underestimates 
T$_*$, mainly for nebulae with high-temperature stars.
The results showing that the stellar temperature can be higher than 
T$_Z$(He II) are consistent with UV data for the central stars of PNs 
(Pottasch et al. 1978; Kaler \& Feibelman 1985).
The relation between the Zanstra temperature ratio ZR and  
T$_Z$(He II) (Fig. 2) can be used to obtain a more accurate  
estimate of the stellar temperature.  
However, for nebulae with very high stellar
temperature and/or small optical depths the theoretical 
results for different stellar temperatures and optical 
depths are crowded and the method is uncertain.  
For these nebulae, a better determination  
of these parameters can be obtained  
from a plot of He II/He I versus T$_Z$(He II) (Figs. 4).  

An important source of uncertainty is related to the kind 
of observations used to determine the stellar temperature.  
Observed line intensities   
do not always refer to the whole nebula since line ratios are usually 
obtained 
from observations with a narrow slit crossing the nebula.  
Furthermore, the nebula may be inhomogeneous and the 
optical depth anisotropic. 
That is, the nebula can be optically thick in some direction but optically 
thin in others. This is a very important point not addressed in this 
paper. Line intensity ratios obtained with a narrow slit, or in a given 
position of a nebula, may not correspond to the ratio for the entire nebula
(Gruenwald, Viegas, \& Brogui\`ere 1997; Gruenwald \& Viegas 1998). 
 
Cases of very high or very low ZR can be explained by  
the coupled effect of optical depth and over- or under-abundance  
of He in the stellar atmosphere, which affects the ionizing spectrum. 
A nebular overabundance of He relative to solar values 
and a low optical thickness explain the very high   
discrepancy shown by some Abell planetaries. 
On the other hand, ZRs less than unity 
are characteristic of optically thick PNs with undersolar He abundance.  
 
Stanghellini et al. (1993) state that since bipolar nebulae have  
T$_Z$(He II) $\sim$ T$_Z$(H), they are thicker than other nebulae.  
However, central stars of bipolar  
nebulae are known to have high temperatures (Corradi \& Schwarz 1995). 
From our results (Figure 2) these  
nebulae are expected to have T$_Z$(He II) $\sim$~ T$_Z$(H), even if they 
are not  
completely optically thick. 
 
Finally, the ``Zanstra wall'' in the  log L - log T diagram 
(Shaw and Kaler 1989) 
is related to the fact that the Zanstra method underestimates the stellar 
temperature and this effect 
is larger for high-temperature stars. Consequently, 
the lack of high-temperature stars (predicted by the evolutionary 
models) in the log L - log T diagram can be easily understood. 

\acknowledgments
We are thankful to M. Peimbert for his valuable suggestions. 
This work was partially supported by FAPESP, CNPq, and PRONEX/FINEP.

\newpage
\figcaption [] {Fractional abundance distribution of H and He ions: 
dependence  
with (a) the radial distance from the 
center, in units of the maximum radius and 
(b) the optical depth in 13.6 eV. 
The figures are labeled by the stellar temperature adopted in the 
models. 
Thick and thin solid lines correspond, respectively, to the fractional 
abundances of H$^0$ and H$^+$ relative to H, while the fractional abundances 
of He$^0$, He$^+$, and He$^{++}$ relative to He are given, respectively, 
by the dotted, dashed, and dot-dashed lines.} 

\figcaption [] {Ratio of Zanstra temperatures vs. 
T$_Z$(He II) for the 
same models as in Fig. 1. Each solid line is labeled by the corresponding 
stellar temperature in units of 1000 K. For each solid line, increasing 
optical depths 
correspond to decreasing ZR. 
The dashed curves connect the results of completely 
optically thick models with different stellar temperatures.
(a) Results characterized by 
$\tau_{13.6}$ = 1 or $\tau_{13.6}$ = 10 or $\tau_{54.4}$ = 1 are indicated,
respectively, by crosses, triangles, and dots; (b) dots represent 
the Zanstra temperatures derived from observations.}

\figcaption [] {Effects of the nebula He abundance and of a departure from 
a blackbody ionization spectrum on the Zanstra temperatures.
Solid lines correspond to the same models as in Fig. 1
(He/H = 0.115). 
Results for a model with T$_*$ = 150,000 K and a higher abundance 
(He/H = 0.200) are indicated by the dot-dashed line. Dotted lines show 
the results for models with a deficit of high-energy photons in the 
ionizing spectrum for various T$_*$.
The dashed lines join the completely optically thick models.} 

\figcaption [] { He II/He I line intensity ratios versus T$_Z$(He II). 
The solid lines
are labeled by the stellar temperature in units of 1000 K. Models for 
$\tau_{13.6}$ = 1 or $\tau_{13.6}$ = 10 or  $\tau_{54.4}$ = 1 are indicated,
respectively, by the dotted, dot-dashed, and dashed lines.} 

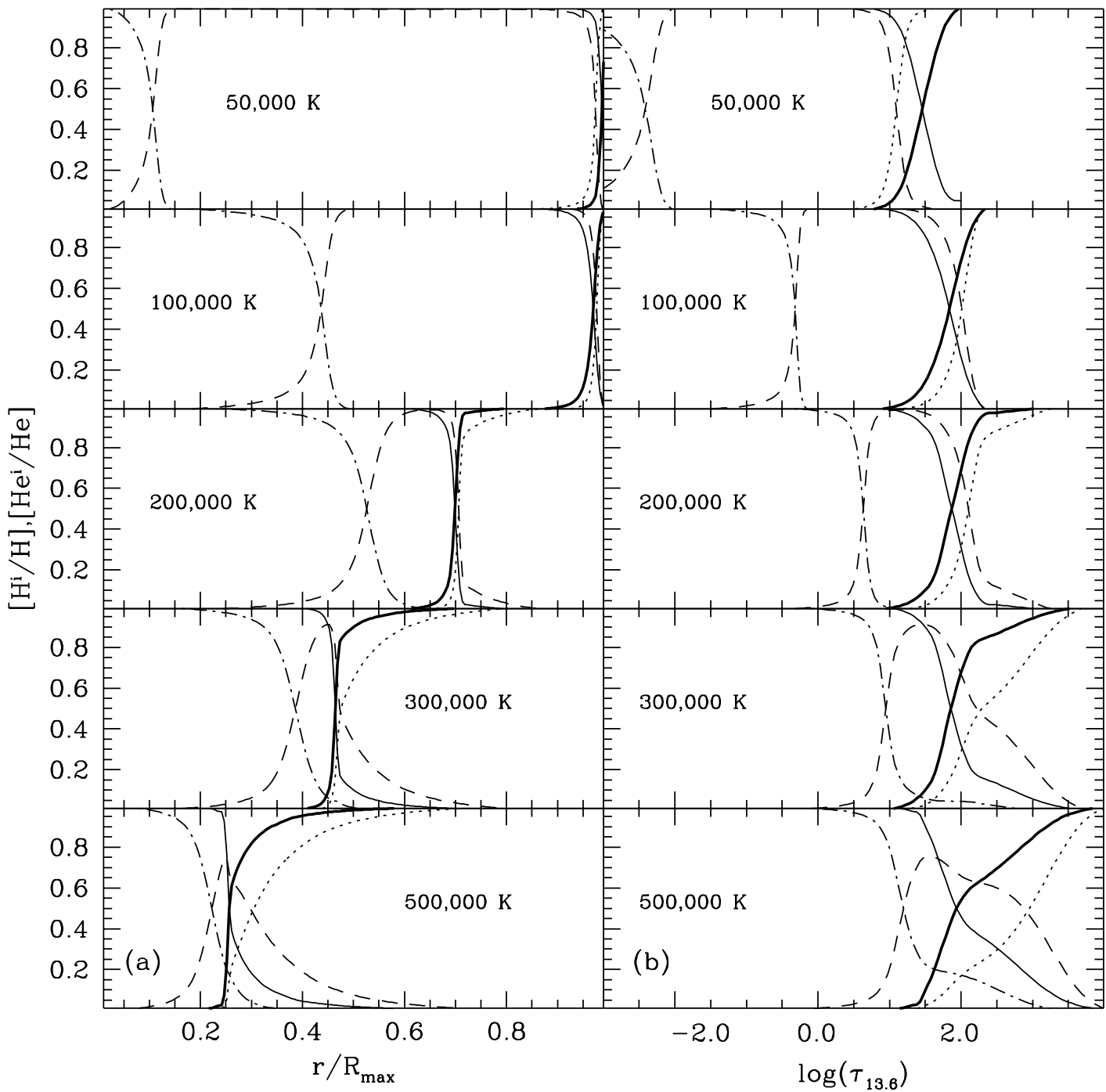
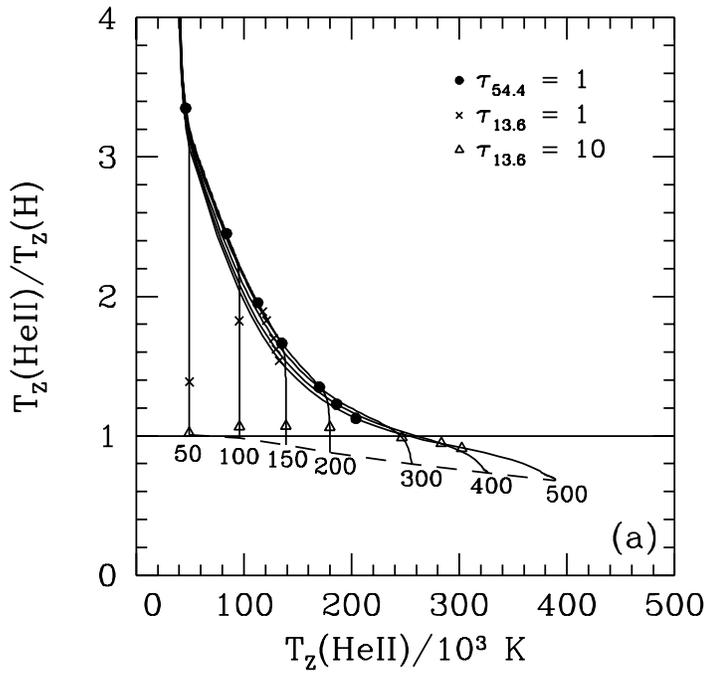
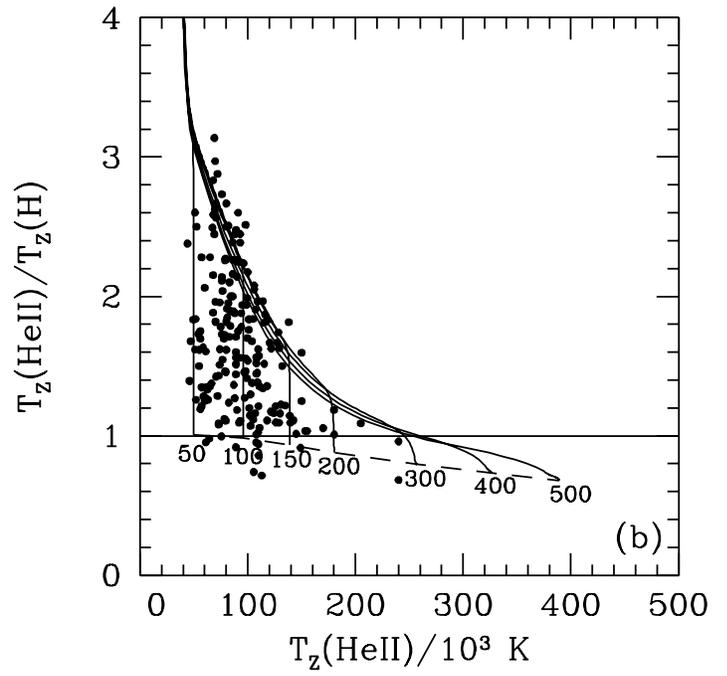
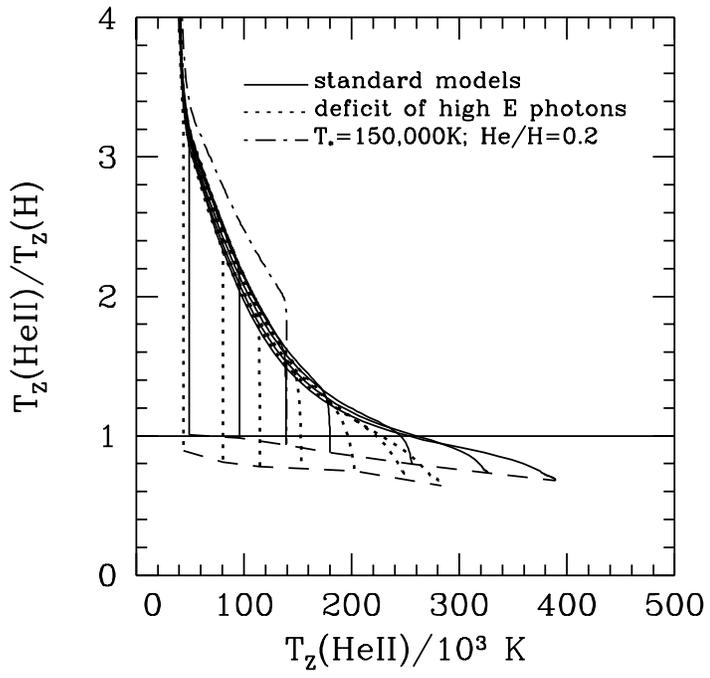
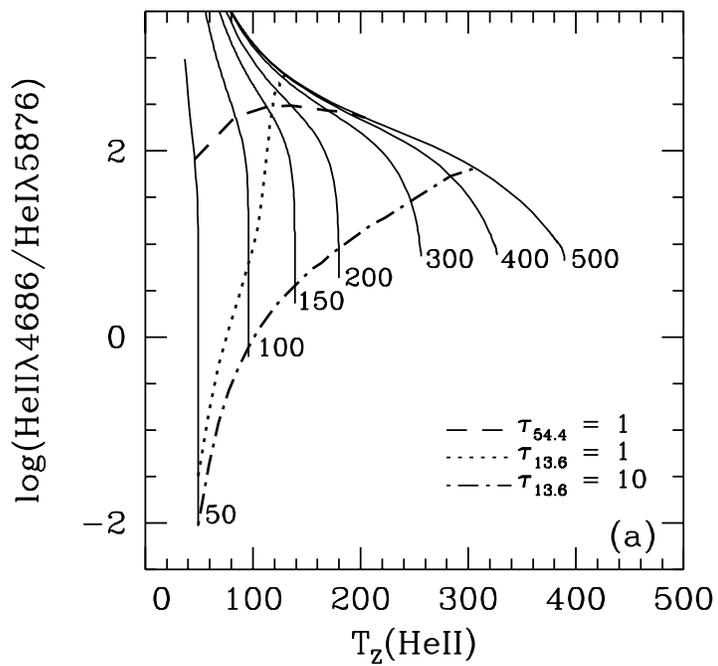
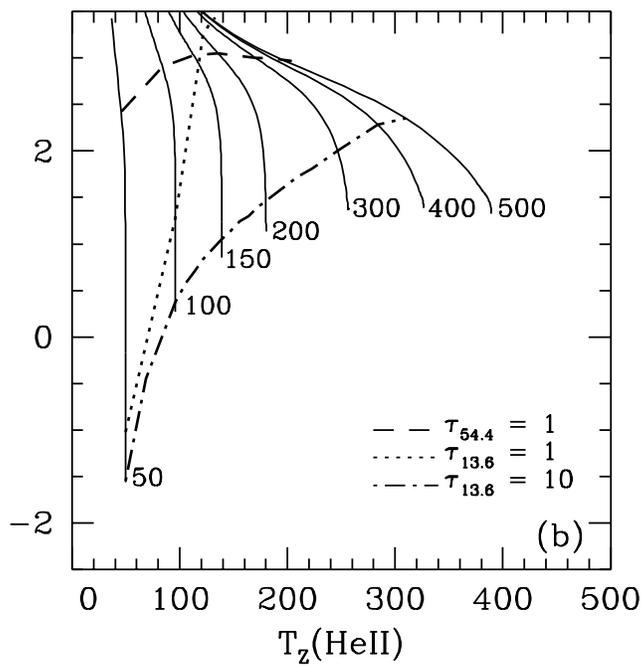
\end{document}